\documentstyle[sprocl,epsfig]{article}

\def\Journal#1#2#3#4{{#1} {\bf #2}, #3 (#4)}

\def\NPB{{\em Nucl. Phys.} B}
\def\PLB{{\em Phys. Lett.}  B}

\def\PRD{{\em Phys. Rev.} D}

\def\be{\begin{equation}}
\def\ee{\end{equation}}
\def\bea{\begin{eqnarray}}
\def\eea{\end{eqnarray}}
\newcommand\place[4]{
  \begin{center}
      \mbox{\psfig{file=#2,width=#1\textwidth}}
      \caption[]{#4}
      \label{#3}
  \end{center}}
\begin{document}

\title{WW PHYSICS AT LEP2\footnote{To appear in the proceedings of the
XXVth International Winter Meeting on Fundamental Physics, Formigal (Spain),
March 1997.}}

\author{R. MIQUEL\footnote{E-mail address: miquel@ecm.ub.es}}

\address{Dept.~ECM and IFAE, Facultat de F\'{\i}sica, Univ.~de Barcelona\\
         Diagonal 647, E-08028 Barcelona (Spain)}

\maketitle\abstracts{
The analyses and results on WW physics obtained in the first year of running of
LEP2 are summarized. The determination of the W mass, both through the 
measurement of the WW production cross section at a center-of-mass energy of 
161~GeV and through the study of the invariant mass distribution of  W decay
products in the 172~GeV run are described first. Preliminary results on
searches for anomalous triple gauge-boson vertices are also presented.}

\section{Introduction}
\label{sect:intro}
The LEP accelerator at CERN is just in the middle of an energy upgrade
that will bring its energy to about 195 GeV in year 1998, with peak 
luminosities 
close to $10^{32}$~cm$^{-2}$s$^{-1}$. 

The main reason to double the energy of the LEP machine is to be able to
produce pairs of W's. Their mass can be accurately measured, providing a new
constraint on the Minimal Standard Model (MSM) fundamental parameters. 
A very precise measurement of the W mass could provide useful information 
on the 
mass of the MSM Higgs boson.
The W 
couplings to the other vector bosons in the electroweak theory (photon and Z),
which come from the non-abelian nature of the gauge group, can be studied with
great detail and constraints on new physics can be derived from their
determination.

%Of course, there are other very intersting physics topics at LEP2. 
%Opening a new energy domain in $e^+e^-$ collisions implies being able to
%search for new particles and phenomena in new regimes. In particular, a wide 
%class of Supersymmetric models
%predict a Higgs boson with a
%mass below 150~GeV. For a wide range of the
%Minimal Supersymmetric Standard Model (MSSM) parameters, this particle
%would be actually accessible at LEP2.

This article summarizes the first results on WW physics
obtained with the 1996 data sample and will analyse also the prospects for
precise measurements of both the W mass and its couplings to Z and photon when
the whole data sample will be collected around year 2000.

The 1996 LEP2 run was splitted into two parts. The first run, between June and
August 1996 collected 11~pb$^{-1}$ of data per experiment at a 
center-of-mass energy $\sqrt{s}=161.3$~GeV, that is 
about 0.5~GeV above the nominal WW production threshold.
The second run, in October and November 1996, was at $\sqrt{s}=172$~GeV and
accumulated 10~pb$^{-1}$.
Final results from the WW analyses for the first set
are available from all four LEP experiments~\cite{aleph_161}--\cite{opal_161}
and will be reviewed here. At the time of writing, only 
preliminary results were available from the
172~GeV run. 
They will also be discussed here, although with less detail, due to
their volatility.

The current LEP2 schedule includes running in years 1997--1999 with 
center-of-mass enegies increasing from 184~GeV in 1997 to 194~GeV in 1998 and
1999. The expected luminosities are around 100~pb$^{-1}$ for 1997 and about
180~pb$^{-1}$ in each of 1998 and 1999. Together with the 21~pb$^{-1}$
collected in 1996, this would bring the total close to 500~pb$^{-1}$, which 
will be taken in the following as ultimate LEP2 luminosity per experiment.

Section~\ref{sect:general} discusses some generalities about W production
and decay, selection and background in $e^+e^-$ collisions. The W mass
determination 
done with 1996 data taken near the WW production threshold
is presented in section~\ref{sect:mass1}. The measurement done with the
1996 data at higher energy (172~GeV)
and the prospects using all data up to 1999 are the subject of 
section~\ref{sect:mass2}. Section~\ref{sect:tgv} contains the
first results and
future capabilities of LEP2 in the measurement of the trilinear gauge-boson 
couplings. Finally, section~\ref{sect:summ} contains a brief summary with the
conclusions of the paper.
\section{WW production and decay}
\label{sect:general}
Pairs of W$^+$s and W$^-$s 
are produced at LEP2 based on the three tree-level diagrams
depicted in fig.~\ref{fig:feyn}. 
%
%\begin{figure}
%\vspace{10cm}
%\caption[a]{Feynmann diagrams for the production of a $W^+W^-$ pair at LEP2.}
%\label{fig:feyn}
%\end{figure}
% 
\begin{figure}
\begin{center}
\place{1.0}{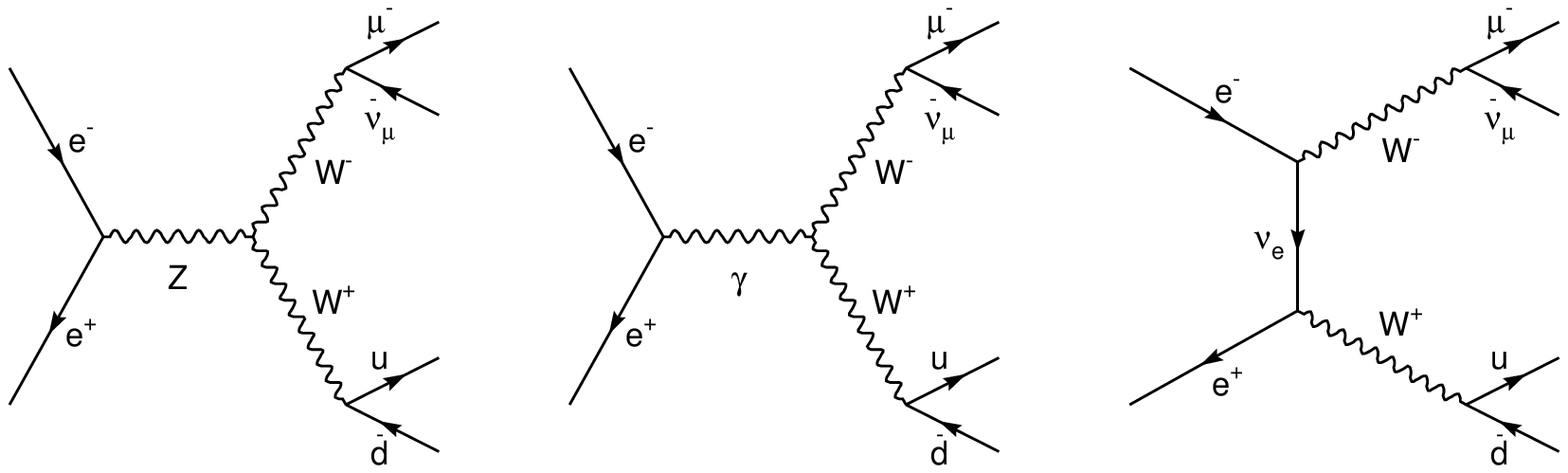}{fig:feyn}{Feynmann diagrams for the production and decay
of a $W^+W^-$ pair at LEP2.}
\end{center}
\end{figure}
The t-channel exchange diagram dominates the resulting amplitude by 
having a factor
$\beta=\left(1-4M_W^2/s\right)^{1/2}$ less. 
Therefore, the Ws tend to be produced
in the forward direction. More so near threshold, where $\beta$ is indeed
very small. The other two diagrams, whose existence is needed in order to
preserve unitarity, include the non-abelian 
couplings between two Ws and either a photon or a Z. Since the t-channel
diagram dominates near threshold, it is clear that the experimental sensitivity
to trilinear gauge-boson couplings will be higher for higher 
center-of-mass energies.

%Figure~\ref{fig:xs}, taken from~\cite{YR_sigw},
%shows the WW production cross section at LEP2 energies in
%the on-shell and off-shell approximations and including 
%initial state radiation.
%
%\begin{figure}
%\vspace{10cm}
%\caption[a]{Total WW production cross section at LEP2 as a function of the
%center-of-mass energy, under different approximations. Taken 
%from~\cite{YR_sigw}.}
%\label{fig:xs}
%\end{figure}
%
%At 161~GeV the total WW cross section is as small as 3.5~pb, while it reaches
%its maximum value of 17~pb around $\sqrt s = 210$~GeV.

Neglecting flavour-mixing, the W boson decays to each pair of available 
up-antidown fermions with the same strength. Therefore, its branching ratios
are 1/3 to lepton and neutrino and 2/3 to quark-antiquark. Because of the
large mass of the top quark, decays to top-bottom pair are not allowed and,
therefore, neglecting bottom-strange and bottom-down mixing, no
$b$ quarks are produced in W decays. 
%This fact is very useful when searching
%for Higgs boson (which tend to decay to $b\bar{b}$ pairs) and trying to
%remove the WW background.

Considering now W
pairs, there are three possible final states:
\begin{itemize}
\item $l l \nu \nu$~(10.5\% 
      of the total WW sample). 
      There are at least two neutrinos (more if $l=\tau$).
      The complete reconstruction of the event is possible only if the value of
      the
      W mass is assumed. These events can be used in the determination of the
      anomalous couplings.
\item $l \nu\ jet\ jet$~(44\%).  
      With one neutrino ($l=e,\mu$, 30\%) 
      the complete
      kinematical reconstruction is possible. These are clean events, with the
      two Ws clearly separated. They can be used in any study presented below.
\item $jet\ jet\ jet\ jet$~(45.5\%). 
      Often with a fifth jet due to gluon
      radiation. These are messy, unclear events, with important
      QCD backgrounds. 
      The task
      of assigning particles to jets and jets to Ws is not easy. While the
      possibility of using the four-jet events for the determination of the W
      mass is by now firmly established, it still remains to be seen whether 
      they can be used effectively in the anomalous couplings analysis.
\end{itemize}

The two main topics of interest in W physics at LEP2 are
the precise determination of the W mass and the study of the non-abelian
couplings between two Ws and Z or $\gamma$. They will be discussed in turn
in the following.
\section{W mass measurement at threshold}
\label{sect:mass1}
The current best determination of $M_W$ comes from the 
CDF~\cite{cdfmw,cdfmw2} and 
D0~\cite{d0mw} experiments 
at the Tevatron, which get  the W mass through a fit to the
``transverse mass'' distribution measured in W leptonic decays from the lepton
momentum and the missing transverse momentum. Their current averaged value 
is~\cite{cdfmw2}
$$
M_W = 80.37 \pm 0.10\,{\mathrm GeV}\,.
$$
This number makes use of all available CDF and D0 data. By the
end of run II,
expected to start in 1999, the error could go down to 30--40~MeV.

An indirect $M_W$ determination can be obtained from the Standard Model fits
to all electroweak data, mainly LEP1 results. Once the free parameters of
the minimal standard model, like $m_t$ or $M_H$ are obtained from the fit,
they can be used to predict any other observable like $M_W$.
The current best value, not using the direct $M_W$ and $m_t$ 
information, is~\cite{lepewwg}
$$
M_W = 80.323 \pm 0.042\,{\mathrm GeV}\, ,
$$
in excellent agreement with the direct measurement. 

A strong 
consistency check of the theory can be obtained by comparing both numbers.
The uncertainty in the indirect determination is going to decrease only 
slightly in the near future (thanks to the continuing $\sin^2\theta_W$ 
measurements at SLAC). Therefore a crucial
goal for the direct $M_W$ determinations is to bring down the error to the
30--40~MeV level, comparable to the error of the indirect measurement.

Two very different methods exist for the measurement of the W mass at LEP2:
\begin{itemize}
\item Measuring the total WW production cross section near the threshold region
allows the determination of the kinematical threshold and, therefore, the W 
mass. The advantadges of this method is that only involves counting events, it
is clean and uses all decay channels. However, it needs data very close to 
threshold, where the cross section is small and, threfore, signal to background
ratios are low and the number of events (for other measurements) is minimal.
\item The direct reconstruction of the invariant mass of hadronic and leptonic
decays uses about 90\%
of the events (all but the ones with two leptonic decays) and
it works equally well at 
any LEP2 energy above about 170~GeV. However, it is very involved, especially
for four-jet events, and it can be sensitive to soft QCD effects, also for the
four-jet channel. 
\end{itemize}
Both methods rely on the knowledge of the beam energy, the second because some
sort of kinematical fit assuming four-momentum conservation is needed in order
to improve the invariant mass resolution. In both cases, the error on the beam
energy, $\Delta E_b$, translates almost directly into an error on the W mass:
$$
\Delta M_W ({\mathrm beam})\simeq \Delta E_b \, .
$$
 
%Near the threshold region, the statistical
%error on the W mass obtained from the 
%determination of the WW production cross section can be written as:
%$$
%\left(\Delta M_W\right)({\mathrm stat}) =
%\left|\frac{dM_W}{d\sigma_{WW}}\right| \Delta\sigma_{WW}({\mathrm stat}) =
%\sqrt{\sigma_{WW}}\left|\frac{dM_W}{d\sigma_{WW}}\right|
%\frac{1}{\sqrt{\epsilon_W{\cal L}}}\, ,
%$$
%where $\epsilon_W$ is the WW selection efficiency. In the last expression,
%the first two terms do not depend on efficiency or luminosity and give us
%the intrinsic
%sensitivity of the measurement. The middle line in fig.~\ref{fig:sensi}
%shows that the minimum of $\sqrt{\sigma_{WW}}\left|dM_W/d\sigma_{WW}\right|$
%is found around $\sqrt s = 2M_W + 0.5$~GeV.
%\begin{figure}
%\vspace{10cm}
%\caption[a]{The middle curve shows the value of 
%$\sqrt{\sigma_{WW}}\left|dM_W/d\sigma_{WW}\right|$ as a function of 
%$\sqrt s -2M_W$.
%%Taken from~\cite{YR_lep2}.
%}
%\label{fig:sensi}
%\end{figure}
%Therefore, this will be the energy
%that will minimize the statistical error on $M_W$. 
%Assuming $M_W\sim 80.3$~GeV, one gets $\sqrt s\sim 161$~GeV as the optimal 
%center-of-mass energy to do the $M_W$ determination near threshold. This was,
%indeed, the energy chosen for LEP2's first run in summer 1996.

The main problem for the measurement of the WW cross section near threshold 
is that the background is much larger than the signal, in particular for the
fully hadronic channel. While the cross section for WW production 
and subsequent
decay to hadrons is about 1.6~pb at 161~GeV, the cross section for hadron 
production through photon and/or Z reaches about 150~pb, or 100 times more.
%If one allows a certain amount of background, 
%$\sigma_{\mathrm bkg}^{\mathrm eff}$, into the final sample, the statistical 
%error on $M_W$ increases from $\Delta M_W$ to 
%$\Delta M_W 
%(1+\sigma_{\mathrm bkg}^{\mathrm eff}/\epsilon_W\sigma_{WW})^{1/2}$. One 
%should try to minimize $\sigma_{\mathrm bkg}^{\mathrm eff}$ while trying to 
%keep $\epsilon_W$ as large as possible.

Studies previous to the start-up of LEP2~\cite{YR_mw} showed that for a
luminosity of 100~pb$^{-1}$, the overall statistical error on $M_W$ could be
of order 134~MeV. Systematic errors were thought to be considerably smaller, 
leading to a total uncertainty around 144~MeV, when considering four
experiments collecting 25~pb$^{-1}$ each.

\subsection{The $WW\to l\nu l\nu$ channel at 161~GeV}
\label{sect:lvlv}
The events in which both Ws decay leptonically are very simple indeed.
%(fig.~\ref{fig:lvlv}).
%% 
%\begin{figure}
%\begin{center}
%\place{0.75}{event_lvlv.eps}{fig:lvlv}{A fully leptonic WW event at 161~GeV
%as seen by the ALEPH detector.}
%\end{center}
%\end{figure}
%% 
All experiments select these events based on their low charge and neutral 
multiplicity, high missing transverse momentum (due to the neutrinos) and high
acoplanarity (since there is more than one invisible particle). The overall 
efficiency varies with experiment, from about 40\%
%in the case of L3
to almost 70\%.
%for ALEPH. 
Differences amongst experiments have to do with their different 
solid angle coverage for track measurements and their diverse capabilities in
tau identification. Indeed, a big fraction of the inefficiency comes from
events were at least one of the Ws decay into tau.

Backgrounds are estimated to be very low, around 50~fb, to be compared with a
signal cross section around 400~fb. The four experiments together
have collected 12
events of this type. Clearly, with this level of statistics, it 
is not susprising that systematic errors are negligibly small, compared to 
the statistical error. 
%Figure~\ref{fig:lvlv_l3} shows the distribution of
%acoplanarity measured by the L3 collaboration. A cut at about 10$^\circ$
%selects very clearly two events from the expected background.
% 
%\begin{figure}
%\begin{center}
%\place{1.0}{eps95_1.eps}{fig:lvlv_l3}{Distribution of acoplanarity as
%used in the selection of the $l\nu l\nu$ final state by the L3 collaboration.}
%\end{center}
%\end{figure}
%
%\begin{figure}
%\vspace{10cm}
%\caption[a]{Distribution of acoplanarity as
%used in the selection of the $l\nu l\nu$ final state by the L3 collaboration.}
%\label{fig:lvlv_l3}
%\end{figure}
%
\subsection{The $WW\to l\nu qq$ channel at 161~GeV}
\label{sect:lvqq}
The selection of $WW\to l\nu qq$ when $l=e,\mu$ is also very simple. 
%One such
%event can be seen in fig.~\ref{fig:lvqq}.
%% 
%\begin{figure}
%\begin{center}
%\place{1.0}{event_lvqq.eps}{fig:lvqq}{A $WW\to\mu\nu qq$ event at 161~GeV
%as seen by the ALEPH detector.}
%\end{center}
%\end{figure}
%% 
All experiments select these events requiring large charged-particle 
multiplicity, large missing energy and momentum and trying to identify an
isolated electron or muon opposite to the direction of the missing momentum.
Efficiencies in this channel range from 70 to 90\%,
the differences coming esentially
from the different lepton coverage of the experiments. Background is very
low, around 20~fb for a signal cross section close to 1.1~pb.

When the lepton is a tau, instead, 
the analysis is substantially more difficult.
The selections differ for each experiment. The most common version selects
events with two broad jets ($q\bar q$), 
one narrow jet (tau) and missing energy.
Efficiencies are substantially lower, 40--55\%,
and background from $q\bar q$ production is rather large 
at about 150~fb, with signal cross section 
around 550~fb. Figure~\ref{fig:tvqq_opal} shows two distributions used by 
the OPAL collaboration
to select $\tau\nu qq$ events, involving the missing energy and the 
direction of the missing momentum.
\begin{figure}
\begin{center}
\place{1.0}{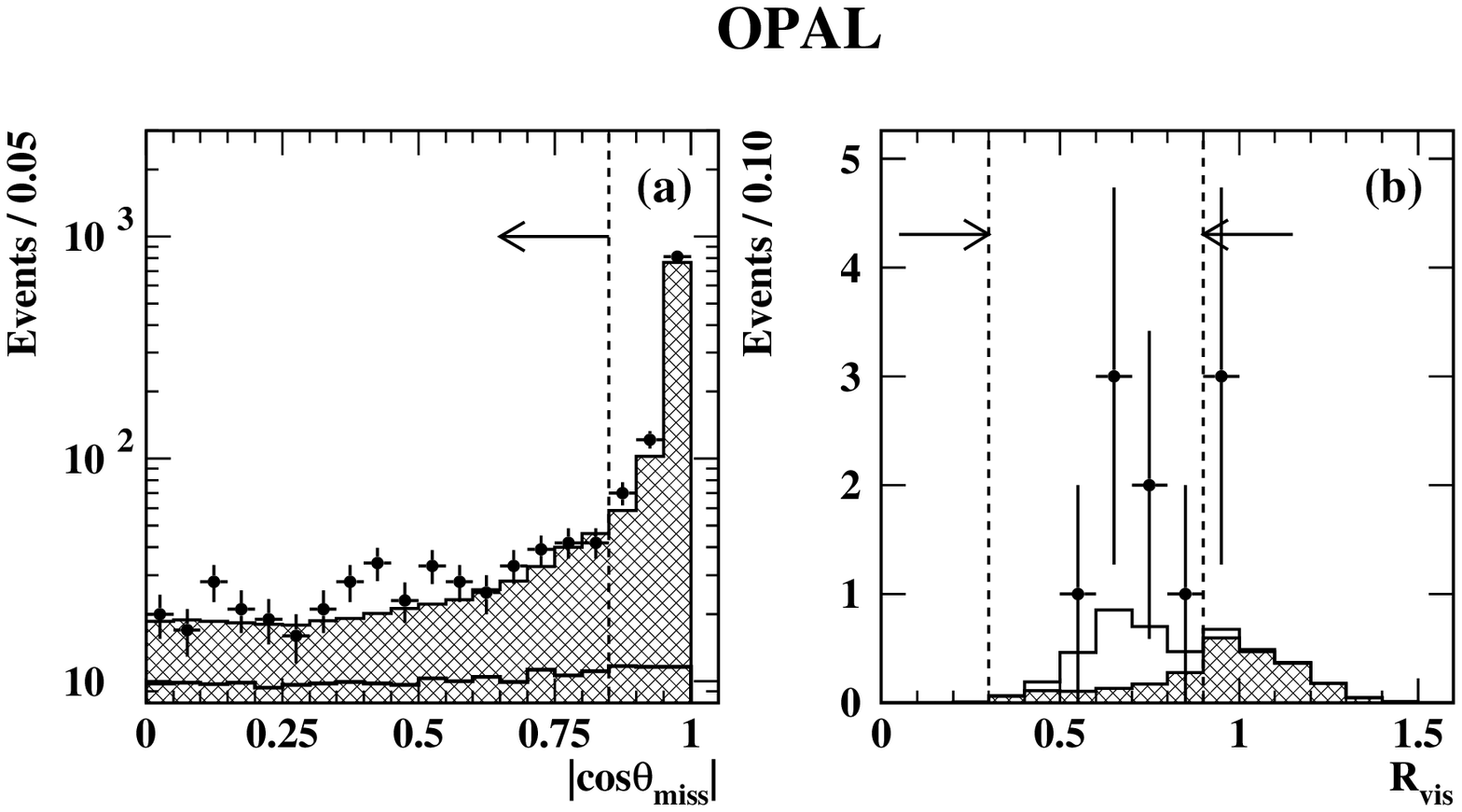}{fig:tvqq_opal}{Distributions of direction of 
missing momentum and of amount of missing energy as
used in the selection of the $\tau\nu qq$ final state by the 
OPAL collaboration. The arrows show the location of the selection cuts}
\end{center}
\end{figure}

The total number of $l\nu qq$ events with $l=e,\mu,\tau$ selected in
the 161~GeV run by the four collaborations is 51. Systematic errors are still
small compared to the statistical errors and are dominated by the subtraction
of the $q\bar q$ background to the $\tau\nu qq$ channel.
\subsection{The $WW\to qqqq$ channel at 161~GeV}
\label{sect:qqqq}
Although WW events in which both Ws decay hadronically at 
$\sqrt s = 161$~GeV do have a characteristic look 
%(fig.~\ref{fig:qqqq}) 
as clear four-jet events with two pairs of back-to-back jets,
their selection is much more involved, due to the huge background 
from $e^+e^-\to q\bar{q}$ production. 
%% 
%\begin{figure}
%\begin{center}
%\place{1.0}{event_qqqq.eps}{fig:qqqq}{A $WW\to qqqq$ event at 161~GeV
%as seen by the ALEPH detector.}
%\end{center}
%\end{figure}
%% 
DELPHI and OPAL have used standard cut
techniques and get efficiencies around 60\%
and signal to background ratios between two and three. L3 and, especially, 
ALEPH have chosen more sophisticated methods based on multivariate analysis.

L3 has used a neural network with 12 variables, which include,
among others:
\begin{itemize}
\item $y_{34}$, the value of the clustering parameter $y_{ij}$ for which the
event changes from having four jets to having three jets;
\item sphericity;
\item the sum and difference between the two reconstucted W masses 
(after choosing one particular pairing among jets);
\item the minimum and maximum jet energies.
\end{itemize}
%Two of the variables can be seen in fig.~\ref{fig:qqqq_L3}, where it is
%obvious that, although they do discriminate somehow between signal and
%background, the discrimination is not such that one could put a cut on any of
%them.
% 
%\begin{figure}
%\begin{center}
%\place{1.0}{eps95_1.eps}{fig:qqqq_L3}{Distributions of two of the 
%variables used in L3's $qqqq$ 
%neural network analysis: $y_{34}$, and the sum of the
%two jet-pair invariant masses, $M_1+M_2$.}
%\end{center}
%\end{figure}
%
%\begin{figure}
%\vspace{10cm}
%\caption[a]{Distributions of two of the 
%variables used in L3's $qqqq$ 
%neural network analysis: $y_{34}$, and the sum of the
%two jet-pair invariant masses, $M_1+M_2$.}
%\label{fig:qqqq_L3}
%\end{figure}
%%
%However, when combined by the neural network, its output, 
%fig.~\ref{fig:qqqq_L3_2}, shows a much clear separation between the WW signal
%and the $q\bar q$ background.
% 
%\begin{figure}
%\begin{center}
%\place{1.0}{l3_nn.eps}{fig:qqqq_L3_2}{Distribution of the output of the
%neural network used by L3 for the $qqqq$ channel.} 
%\end{center}
%\end{figure}
%
%\begin{figure}
%\vspace{10cm}
%\caption[a]{Distribution of the output of the
%neural network used by L3 for the $qqqq$ channel.}
%\label{fig:qqqq_L3_2}
%\end{figure}
%

The ALEPH collaboration has investigated several multivariate techniques for
the $qqqq$ selection, which have been combined at the end. In all the analyses,
loose preselections are applied to the data to get rid of clear 
$e^+e^- \to \gamma Z \to \gamma q \bar{q}$ radiative return events, two-photon
initiated events and two-jet like events. After preselection, the 
efficiency is still very high, about 90\%,
but the signal to background ratio is still only around 0.1--0.3.

Then, signal is separated from background by using simultaneously several
variables, most of them close to those used by L3 with the addition of two
more discriminating quantities: the value of the four-jet QCD matrix element
squared evaluated with each event kinematic variables, and the sum of 
the transverse momentum of
all particles with respect to their nearest jet. 
%Some of the variables are
%shown in fig.~\ref{fig:qqqq_ALEPH}.
%% 
%\begin{figure}
%\begin{center}
%\place{1.0}{aleph_qqqq1.eps}{fig:qqqq_ALEPH}{Distributions of some of the 
%variables used in ALEPH's $qqqq$ 
%analysis. The points represent the data; the dark shaded histograms show the
%WW expectation; the light shaded histogram show the background expectation 
%and the histogram line is the sum of both.}
%\end{center}
%\end{figure}
%

Four different statistical techniques have been used to combine the information
on all those variables: a linear discriminant variable, a rarity variable,
a weight technique and, finally a neural network. Detailed information on all
four techniques can be found in ref.~\cite{aleph_161}. 
%Figure~\ref{fig:qqqq_ALEPH_2} shows the outputs of the four techniques, in
%which it is clear the separation between signal and background achieved.
%% 
%\begin{figure}
%\begin{center}
%\place{1.0}{aleph_qqqq2.eps}{fig:qqqq_ALEPH_2}
%{Distribution of the output of the four multivariate analyses
%used by ALEPH for the $qqqq$ channel. 
%The points represent the data; the dark shaded histograms show the
%WW expectation; the light shaded histogram show the background expectation 
%and the histogram line is the sum of both.} 
%\end{center}
%\end{figure}
%%
The results of the four analyses are then combined taking into
account the correlations
between them, which are found to be around 0.5--0.6, and a cross section for
the $qqqq$ channels is obtained, $\sigma_{qqqq} = 1.80 \pm 0.50 \pm 0.19$~pb.
The systematic error is larger than for the other two channels and has 
several components dealing with 
detector modeling, background subtraction, etc. 
\vspace*{1cm}

Finally, the results coming from the three channels are combined using a 
maximum likelihood method with inputs given by the number of observed events in
the purely leptonic and the lepton-hadron channel and the cross section
measured in the fully hadronic channel. 
%The Standard Model W branching ratios
%are assumed, and the resulting values of $\sigma_{WW}$, understood as 
%{\tt CC03} cross section, are shown in fig.~\ref{fig:xs_161}.
%% 
%\begin{figure}
%\begin{center}
%\place{1.0}{lep_4_sigw_161.eps}{fig:xs_161}
%{The WW cross sections at $\sqrt s= 161$~GeV measured by the four experiments
%and their combination.}
%\end{center}
%\end{figure}
%
The combination of the WW cross sections measured by the four experiments,
assuming the Standard Model W branching ratios,
gives the final result
$$
\sigma_{WW}(161.3\,\mbox{\rm GeV}) = 
\left( 3.69 \pm 0.45 \right)\mbox{\rm  pb}\, .
$$
The error is dominated by statistics. The systematic error (0.15~pb, included
in the 0.45~pb) is dominated by the contribution from the $qqqq$ channel.

From the total WW cross section at threshold, the W mass is obtained by
using a Standard Model calculation that relates the WW cross section near
threshold to its mass~\cite{gentle}. Most of the W mass sensitivity comes 
really from phase space restrictions and is, therefore, independent of many of
the Standard Model assumptions. The resulting W mass obtained from the
threshold cross section measurement at 161~GeV is:
$$
M_W = \left( 80.40 ^{+0.22}_{-0.21} \pm 0.03 \right) \mbox{\rm  GeV}\, ,
$$
where the last error reflects the current uncertainty on the LEP beam energy.
%Figure~\ref{fig:mw_161} shows the four W masses obtained by the four 
%experiments, while 
Figure~\ref{fig:xsvsmw} shows the relationship between the
measured cross section and mass.
%% 
%\begin{figure}
%\begin{center}
%\place{1.0}{lep_4_mw_161.eps}{fig:mw_161}
%{The W mass measured by the four experiments
%and the combined value.}
%\end{center}
%\end{figure}
%%
% 
\begin{figure}
\begin{center}
\place{1.0}{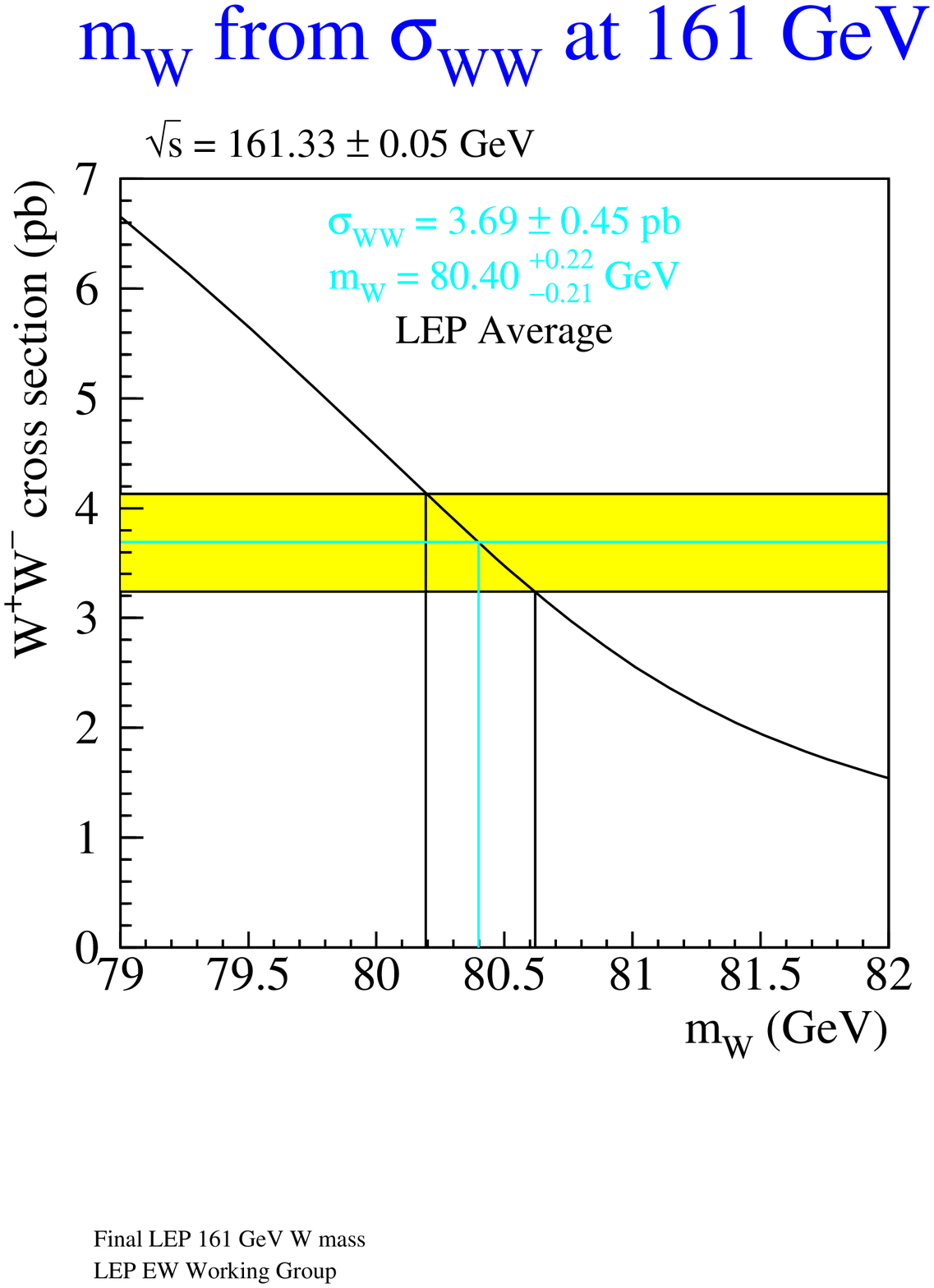}{fig:xsvsmw}
{WW cross section at $\sqrt s= 161.3$~GeV as a function of W mass. The band
represents the experimental measurement.}
\end{center}
\end{figure}
\section{W mass measurement above threshold}
\label{sect:mass2}
Above the WW production threshold, the WW cross sections is no longer very
sensitive to the W mass. The best method to measure the W mass at energies
above 170~GeV is to measure directly the invariant mass distribution of
the W decay products, $M_{jj}$ and $M_{l\nu}$, in $qqqq$ and $l\nu qq$ events.
In the semileptonic events, the momentum of the neutrino can be obtained
imposing four-momentum conservation.
The fully leptonic events cannot be used because of the two missing neutrinos.

Once the invariant mass distribution is measured, a fit to an appropriate
function that will depend on the W mass will give the best estimate for
this parameter, which will be approximately equal to the mean of the invariant
mass distribution.

The detector resolution being far too poor to obtain a reasonably narrow
invariant mass distribution, energy and momentum conservation have to be 
imposed in all channels to improve the resolution. Perfect resolution
would result in a distribution close to a Breit-Wigner with the width 
given by the W decay width.

The intrinsic resolution of the method does not change very much with the
center-of-mass energy, and, therefore, the method works best at the point
with highest cross section, which for the LEP2 range is the point with
highest energy, around 194~GeV. However, the method works similarly well
at all energies above about 170~GeV.

Since the WW cross section increases sharply when moving away from threshold,
while the $q\bar q$ cross section decreases slightly, the background problem
in the fully
hadronic channel discussed in the previous section is much less of a 
problem already at 172~GeV.

However, for this same channel, $qqqq$, other difficulties arise in the jet
finding process and in the jet pairing into Ws, for which there are three
possible combinations that will produce a total of six invariant masses per
event, of which only two have anything to do with the W mass.

%Several 
%techniques have been tried in order to improve the detector resolution:
%\begin{itemize}
%\item In the $l\nu qq$ channel, 
%a full kinematical fit allows to reconstruct the
%full event kinematics. Imposing four-momentum conservation, there are four
%equations and three unknowns (the neutrino momentum), so it is a 1-C 
%(one-constraint) fit.
%\item In the $qqqq$, one can try to use a simple rescaling algorithm, which 
%only changes the jet energies but not their angles. Or one can go for a 4-C
%kinematical fit imposing four momentum conservation. Or 
%even more sophisticated
%approaches like a 5-C fit imposing that the two W masses are equal (within the
%limits given by the W width), ir order to take into account the kinematical
%anticorrelation between the two W masses when being relatively close to the WW
%threshold.
%\end{itemize}

An example of what the analysis chain could look like in the four-jet channel
will clearly show the fact that this is not, indeed, a simple analysis:
\begin{enumerate}
\item Event selection: based on large number of tracks or calorimeter clusters,
large visible energy and momentum balance. Efficiencies in excess of 70\%
can be obtained with low levels of background.
\item Jet reconstruction: using the Durham or Jade algorithms with a value
for $y_{cut}$ that has to be optimized for maximum jet-pair mass resolution.
\item Jet assignment to Ws: decide which jet goes to which W based on some
$\chi^2$ value, or rough mass estimate. Or just take all 
possible combinations.
\item Kinematical fit: fit all jet energies and angles imposing four-momentum 
conservation and,
optionally, an additional constraint on the difference of the two W masses.
If this is not down, one will obtain two masses per event, which will, in
general, be highly anticorrelated.
\item W mass fit: fit the resulting invariant mass distribution with a formula
that has to represent a Breit-Wigner with threshold effects and include 
initial state radiation, biases due to the 
method, correlations if two masses per
event are used, etc.
\end{enumerate}

Based on analyses of this kind made on Monte Carlo events, the authors of
ref.~\cite{YR_mw} concluded that the statistical error on $M_W$ 
that could be expected after 500~pb$^{-1}$ would be around 75~MeV for each
of the channels, $l\nu qq$ and $qqqq$, increasing slightly when moving from
a center-of-mass energy of 175~GeV to 192~GeV.

Systematic errors were also considered. Due to the use of four-momentum 
conservation, the beam energy uncertainty translates directly into an
uncertainty on the W mass. Experimental systematical errors were evaluated as
affecting $M_W$ at the level of 20~MeV or so, while theoretical uncertainties 
in the initial state radiation calculation were estimated as 10~MeV.
Putting all four
experiments together, the expected final error, for a luminosity per experiment
of 500~pb$^{-1}$ was estimated to be 45~MeV for the $qqqq$ channel, 44~MeV
for the $l\nu qq$ channel, and 34~MeV when combining both, roughly independent
of the center-of-mas energy in the range between 170~GeV and 195~GeV.

A possibly important source of systematic error not included in the above 
estimate can come from soft QCD effects. The exchange of soft gluons between
quarks from different W decays can lead to substantial momentum transfer
between the two Ws in a fully hadronic event. This, in turn, would modify
the invariant mass distributions of the two Ws. The effect has been called
``color reconnection'', 
because in an extreme case, can lead to the formation of
a color singlet with a quark and an antiquark coming from different Ws.

Several models have been built to try to assess the size of the effect on 
$M_W$~\cite{color}. The estimates range from esentially no effect, to shifts
of order 50--70~MeV. These are large numbers but,
on the other hand, the effect is small enough so that
it can be very difficult to measure it in any other distribution available
at LEP2. 

Bose-Einstein correlation effects between identical particles 
(for example, pions) from the decay of the two Ws, can also lead to momentum
exchange between the two Ws. For the moment, to take all these possible shifts
into account,
a systematic error of order 50~MeV
should probably be added to the error estimate 
for $M_W$ obtained from direct reconstruction in the $qqqq$ channel.
\subsection{The $WW\to l\nu qq$ channel at 172~GeV}
The selection of semileptonic events at $\sqrt s  = 172$~GeV proceeds in a 
similar way as at 161~GeV. The efficiencies are slightly higher, 70--80\%,
and the background lower.

The event kinematics is reconstructed using a 1-C kinematic fit, as explained
above, or, in some experiments, a 2-C fit, in which the two invariant masses
are taken as equal: $m_{l\nu} = m_{jj}$. The resulting invariant mass 
distribution is then fitted using a simple Breit-Wigner formula. This
introduces a shift on $M_W$ of order 100~MeV, which is corrected for 
by using
Monte Carlo generated events.

%Figure~\ref{fig:lvlv_l3_172} shows the invariant mass distribution measured
%by the L3 collaboration and the comparison with the Monte Carlo prediction.
%% 
%%\begin{figure}
%\begin{center}
%\place{1.0}{eps95_1.eps}{fig:lvlv_l3_172}
%{Average W mass in $l\nu qq$ events at $\sqrt s = 172$~GeV
%as measured by the L3 collaboration (preliminary).}
%\end{center}
%\end{figure}
%
%\begin{figure}
%\vspace{10cm}
%\caption[a]{Average W mass in $l\nu qq$ events at $\sqrt s = 172$~GeV
%as measured by the L3 collaboration (preliminary).}
%\label{fig:lvlv_l3_172}
%\end{figure}
%%
%At this point it should be mentioned that all the results referring to 172~GeV
%data are, at the time of writing, still preliminary for all LEP experiments.
%
\subsection{The $WW\to qqqq$ channel at 172~GeV}
The selection in the fully hadronic channel 
at 172~GeV is much simpler that at 161~GeV
because of the much larger ratio of signal to background. All experiments 
have chosen to do a standard cut selection, with efficiencies around 70--80\%.

Most experiments then use a 5-C kinematical fit to improve their mass
resolution, imposing four-momentum conservation and equality of the two
W masses. The assigment of jets to pairs is done either according to some
$\chi^2$ obtained from the 5-C fit or according to the distance of the
invariant mass obtained to some reference value, although this last technique
probably introduces some bias in the result.

Finally the fit is made with a Breit-Wigner formula, in some cases including
a polynomial background. One of such fits, from DELPHI, can be seen in
fig.~\ref{fig:qqqq_de_172}.
\begin{figure}
\begin{center}
\place{1.0}{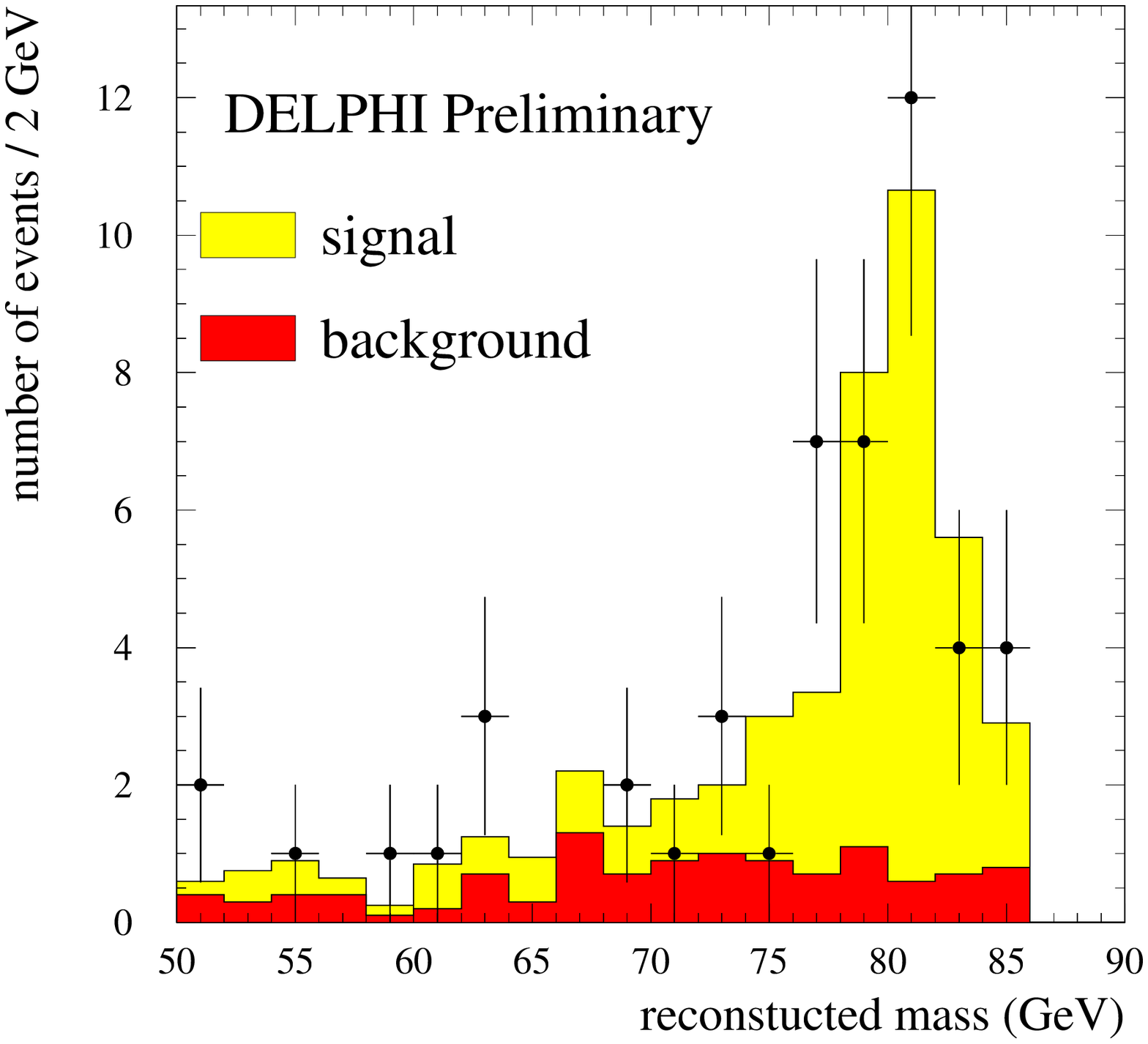}{fig:qqqq_de_172}
{Average W mass in $qqqq$ events at $\sqrt s = 172$~GeV
as measured by the DELPHI collaboration (preliminary).
The points represent data, while the shaded histograms are expectations from
WW signal and background.}
\end{center}
\end{figure}

There are still quite a number of issues that have to be better understood in
all analyses: the optimization of the jet pairing, the use of a
fit formula more complete than a simple Breit-Wigner, the problem of the
soft QCD corrections... The 1997 data will both
allow making more studies to solve
these issues and will bring a decrease in statistical error that will indeed
make all these points very relevant.

For the moment, the combination of the four
preliminary W mass measurements with the direct 
reconstruction method at 172~GeV (including both semileptonic and totally 
hadronic events) gives
%can be seen in fig.~\ref{fig:mw_172}. 
%% 
%\begin{figure}
%\begin{center}
%\place{1.0}{lep_4_mw_172.eps}{fig:mw_172}
%{W mass determinations by the four experiments at 172~GeV (preliminary).}
%\end{center}
%\end{figure}
%%
%The four determinations agree very well and their combination gives
\begin{eqnarray}
M_W &=& \left(80.37\pm 0.18_{exp}\pm 0.05_{theo} \pm 0.03_{beam}\right)
                                  \,{\mathrm GeV}\nonumber\\
    &=& \left(80.37\pm 0.19\right)\,{\mathrm GeV} \, .\nonumber
\end{eqnarray}
%as the preliminary result for the LEP2 $M_W$ determination using direct 
%reconstruction. 
The dominant error includes statistical and purely experimental
errors, the second one is the estimate of the uncertainty due to soft QCD
effects and the third comes from the beam energy uncertainty.

This result can be combined with the final $M_W$ determination
using the WW production cross section measured at 161~GeV to give the 
preliminary W mass result from the first year of running of LEP2:
$$
M_W = \left(80.38 \pm 0.14\right)\,{\mathrm GeV}\, .
$$
This value is compared in fig.~\ref{fig:mwcompa} with the other measurements
of $M_W$, both direct at the Tevatron and indirect at LEP1 and SLD. 
\begin{figure}
\begin{center}
\place{1.0}{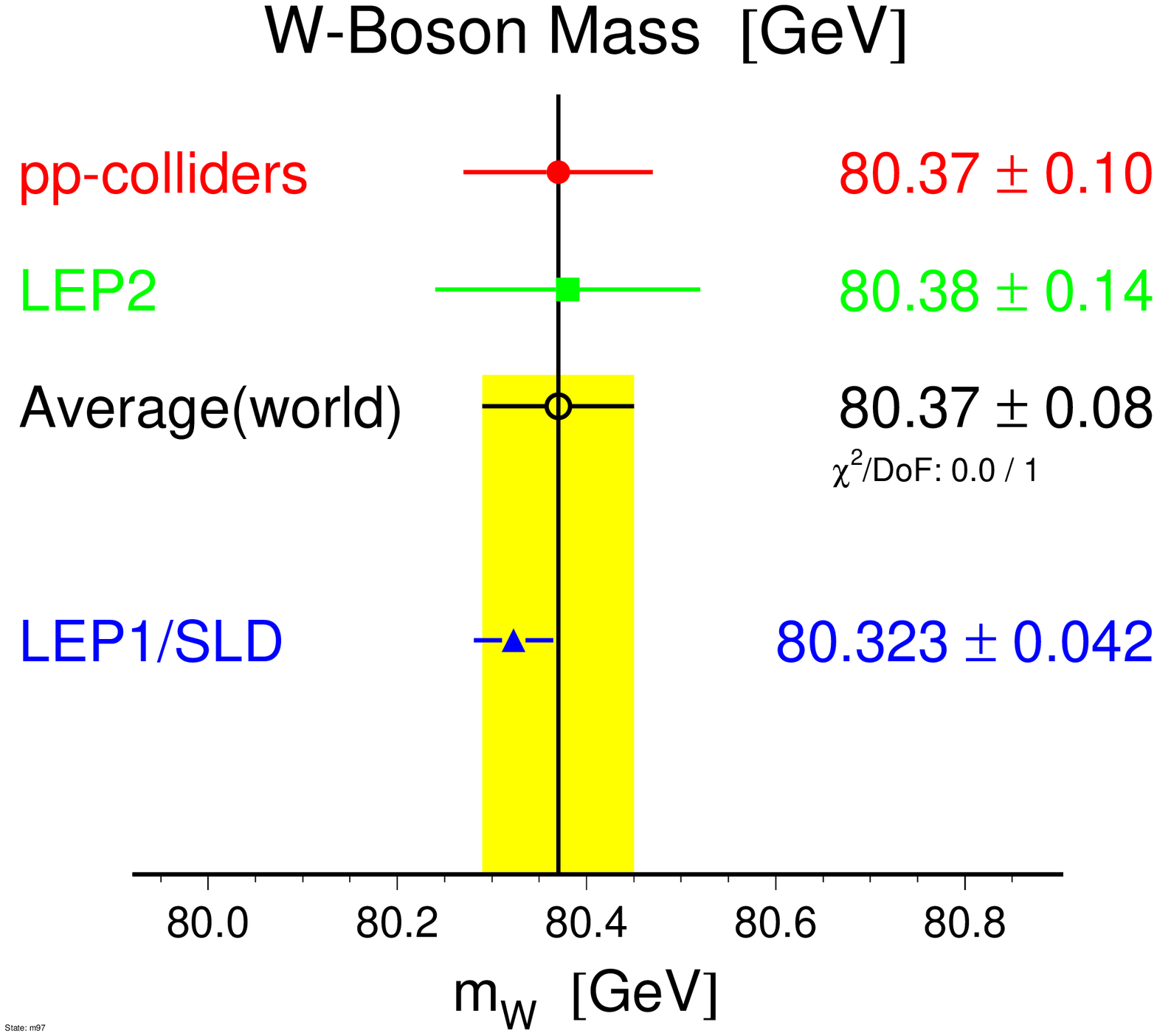}{fig:mwcompa}
{Comparison of the direct W mass measurements at the Tevatron and LEP2 and 
the indirect determination using LEP1 and SLD data.}
\end{center}
\end{figure}
The agreement of all
measurements is perfect. The current LEP2 error is already quite good 
but still much larger than the 40~MeV uncertainty
in the indirect measurement. To get a direct determination of $M_W$ with this
sort of accuracy and to compare it against the indirect measurement, which
assumes the Minimal Standard Model, is the next challenge for both LEP2 and
the Tevatron.
%in their run II, due to start in 1999.
%
%\vspace*{5mm}
%
%The WW cross section has also been measured at 172~GeV. Since it is larger, 
%there are more events, and also they are more easily selected. Therefore,
%it can be measured more accurately than at 161~GeV. However, its sensitivity
%to the W mass is very small. The preliminary combination of the four 
%experiments cross section is
%$$
%\sigma_{WW}(172\, \mbox{\rm GeV}) = 
%\left( 12.05 \pm 0.73 \right)\mbox{\rm  pb}\, .
%$$
%%
\section{Triple gauge boson vertices}
\label{sect:tgv}
Measuring directly the strengh of the non-abelian coupling between three vector
bosons is an important test of the validity of the Minimal Standard Model.
Direct measurements at the Tevatron using $W\gamma$ and $WZ$ production are
not very precise, while
some existing very stringent bounds on deviations from the
MSM predictions obtained using LEP1 data are both model dependent and not
comprehensive.

The most general Lagrangian involving interactions of two Ws and a photon or Z
which is Lorentz invariant and preserves U(1) gauge invariance and parity and
charge conjugation in the electromagnetic sector can be written as:
\begin{eqnarray}
{\cal L} = &-& ie A_\mu \left(W^{-\mu\nu} W^+_\nu - W^{+\mu\nu} W^-_\nu \right)
\nonumber \\
&-& ie \kappa_\gamma F_{\mu\nu} W^{+\mu} W^{-\nu}
\nonumber \\
&-& ie g_{ZWW} Z_\mu \left(W^{-\mu\nu} W^+_\nu - W^{+\mu\nu} W^-_\nu \right)
\nonumber \\
&-& ie \cot\theta_W\kappa_Z  Z_{\mu\nu} W^{+\mu} W^{-\nu}
\nonumber \\
&+& ie \frac{\lambda_\gamma}{M_W^2}  
         F^{\nu\lambda} W^-_{\lambda\mu} W^{+\mu}_\nu
\nonumber \\
&+& ie \cot\theta_W \frac{\lambda_Z}{M_W^2}  
         Z^{\nu\lambda} W^-_{\lambda\mu} W^{+\mu}_\nu
\nonumber \\
&+& e\frac{z}{M_W^2} \partial_\alpha \hat{Z}_{\rho\sigma}
\left(
\partial^\rho W^{-\sigma}W^{+\alpha} - \partial^\rho W^{-\alpha}W^{+\sigma} +
\partial^\rho W^{+\sigma}W^{-\alpha} - \partial^\rho W^{+\alpha}W^{-\sigma}
\right)
\nonumber \\
&+&ie\cot\theta_W \kappa'_Z \hat{Z}_{\mu\nu} W^{+\mu}W^{-\nu}
\nonumber \\
&+&ie\cot\theta_W \lambda'_Z \hat{Z}^{\nu\lambda} W^+_{\lambda\mu}W^{-\mu}_\nu
\nonumber \\
&+&ie\cot\theta_W \kappa''_Z
         \left(\partial^\mu Z^\nu + \partial^\nu Z^\mu\right)W^+_\mu W^-_\nu
\nonumber \, ,
\end{eqnarray}
where 
\begin{eqnarray}
V^{\mu\nu} &=& \partial^\mu V^\nu -\partial^\nu V^\mu \nonumber \\
\hat{V}^{\mu\nu} &=& \frac12\epsilon^{\mu\nu\rho\sigma} V_{\rho\sigma}\nonumber
\end{eqnarray}
for $V=F,W,Z$.

There are ten terms in total, with nine free parameters, the fixed one
being e, the W charge. The first six terms conserve both $P$ and $C$
separately. They include the (electromagnetic and weak)
anomalous dipole magnetic moment and  
quadrupole electric moment of the W, and the point-like coupling of two
Ws to one Z. The anapole term, the one with the $z$ parameter, 
violates both $P$ and $C$ while conserving $CP$. The last three terms 
violate $CP$. In the MSM, the only non zero parameters are $\kappa_\gamma$
and $\kappa_Z$, which both are equal to 1, and $g_{ZWW}$, which 
is $\cot\theta_W$ in the MSM.

Normally one assumes $C$ and $P$ invariance of the whole lagrangian, which
eliminates the last four terms and leaves five free parameters. The next
step in simplification consists in imposing $SU(2)_L\times U(1)_Y$
invariance to the whole lagrangian. Then one can use the effective lagrangian
approach and expand the lagrangian in
invers powers of the square of the scale for new physics. The numerator is 
given by the combinations of the available operators which have the right
dimensions and which are gauge-invariant. 
If one makes use of the light Higgs field to construct the operators and
furthermore the expansion is terminated at the first
non-trivial term (that is, $M_W^2/\Lambda^2_{NP}$ 
is assummed to be small), it is
found that the five non-zero parameters left can be expressed in terms of only
three independent parameters~\cite{YR_tgv}:
\begin{eqnarray}
g_{ZZW} - \cot\theta_W &=& \frac{\alpha_{W\Phi}}{\sin\theta_W\cos\theta_W}
\nonumber \\
\kappa_\gamma - 1 &=& \cot^2\theta_W \left(\kappa_Z-1\right) + 
                      \cot\theta_W \left(g_{ZWW}-\cot\theta_W\right)
\nonumber \\
                  &=& \alpha_{W\Phi} + \alpha_{B\Phi}
\nonumber \\
\lambda_\gamma = \lambda_Z &=& \alpha_W
\nonumber \, .
\end{eqnarray}
Therefore, there are only three free parametes, which can be chosen
as $g_{ZWW}$, $\kappa_\gamma$, $\kappa_Z$ or 
as $\alpha_{W\Phi}$, $\alpha_{B\Phi}$, $\alpha_W$. 
This is the approach that has been followed by the LEP experiments.

If no light Higgs is assumed, then $\lambda_\gamma$ and $\lambda_Z$ are both
of order $M_W^4/\Lambda^4_{NP}$ and one is left, again, with only three free
parameters, $g_{ZWW}, \kappa_\gamma, \kappa_Z$, which are reduced to only two
if global $SU(2)$ symmetry is assumed in the limit $g'\to 0$.

The present direct limits to $|\kappa_\gamma - 1|$ and $|\lambda_\gamma|$
from the Tevatron experiments stand both at around 0.5. Limits on the 
parameters $\alpha_{W\Phi},\alpha_{B\Phi}, \alpha_W$ from LEP1 loop effects
are also in the same range 0.1--1. However, other terms which appear in the 
expansion of the lagrangian and which do not contribute to three-point
functions (vertices) but do contribute to two-point functions (or vacuum
polarizations, which are very well measured at LEP1), like $\alpha_{BW}$,
are strongly constrained
by LEP1 measurements to be below 0.01. It has been argued~\cite{DeRujula}
that, naturally, most models for new physics would tend to predict
$$
\alpha_{BW}\sim\alpha_{W\Phi}\sim\alpha_{B\Phi}\sim\alpha_W \, ,
$$ 
and that, therefore, only effects of order $10^{-2}$ 
could be visible at LEP2. While
this looks indeed a natural assumption, a truly modent-independent study
of the trilinear couplings cannot be based on these assumptions, which, on top,
could be triggered by some peculiar cancellation being at work 
for $\alpha_{BW}$.

The experimental information that can 
be used for the measurement of the couplings
is all contained in the five-fold differential 
distribution
$$
\frac{d^5\sigma}{d\cos\theta_W d\cos\theta^*_+ d\cos\Theta^*_-
d\phi^*_+ d\phi^*_-} \, ,
$$
where $\theta_W$ is the $W$ production angle and $\theta^*_{+(-)}$
and $\phi^*_{+(-)}$ are
the decay angles of the $W^{+(-)}$ in its rest frame. Not all information is 
available for all W decay channels. In the fully hadronic
channel, since it is very
difficult to know which jet is a quark and which is an antiquark, which is 
the $W^+$ and which is the $W^-$, one only has access to folded distributions
of the five angles. In the fully leptonic channel, the kinematics can be solved
by assuming a certain value for the W mass. However one obtains two different
solutions for the five angles. Finally, no problem arises in the $l\nu qq$
channel.

The method for obtaining the values of the trilinear couplings can be
summarized as follows:
\begin{itemize}
\item Select $W^+W^-$ events using standard cut techniques.
\item Use a kinematical fit to improve angular reconstruction.
\item For the fully hadronic channel, choose jet pairing to form Ws.
\item Get the angles $\theta_W,\theta^*_+,\theta^*_-,\phi^*_+,\phi^*_-$.
\item Perform a maximum likelihood fit to the five-dimensional distribution,
having as free parameters one, two or three of the $\alpha_i$ variables
introduced above.
\end{itemize}

Monte Carlo studies show~\cite{YR_tgv} that in fits with only one free
parameter, $\alpha_{W\Phi}$ (the other two are set to zero), limits around
0.05--0.10 can be obtained, dependending on the center-of-mass energy,
with luminosity around 500~pb$^{-1}$.
The sensitivity increases substantially at $\sqrt s = 190$~GeV with respect 
to $\sqrt s = 176$~GeV. Furthermore, the semileptonic channel performs 
substantially better than the fully hadronic channel, and both do better than
the fully leptonic one, because of its reduced branching fraction. 
Systematic uncertainties are expected to be substantially smaller than the
statistical error, even with the full LEP2 data sample.

%When two parameters are left free, the correlation between them degrades the
%resolution substantially, as can be seen in fig.~\ref{fig:aw_awf}.
% 
%\begin{figure}
%\begin{center}
%\place{1.0}{eps95_1.eps}{fig:aw_awf}
%{Monte Carlo study on the sensitivity of a simultaneous fit 
%to $\alpha_{B\Phi}$ and $\alpha_{W\Phi}$ (a--d) and
%to $\alpha_W$ and $\alpha_{W\Phi}$ (e--h). Figures a,b,e,f correspond to 
%the $l\nu qq$ channel, while c,d,g,h show the result for the $qqqq$ chanel.
%All curves show 95\%
%confidence level limits. From ref.~\cite{YR_lep2}.} 
%\end{center}
%\end{figure}
%
%\begin{figure}
%\vspace{14cm}
%\caption[a]{Monte Carlo study on the sensitivity of a simultaneous fit 
%to $\alpha_{B\Phi}$ and $\alpha_{W\Phi}$ (a--d) and
%to $\alpha_W$ and $\alpha_{W\Phi}$ (e--h). Figures a,b,e,f correspond to 
%the $l\nu qq$ channel, while c,d,g,h show the result for the $qqqq$ chanel.
%All curves show 95\%
%confidence level limits. From ref.~\cite{YR_tgv}.}
%\label{fig:aw_awf}
%\end{figure}
%

With the small WW data sample collected at 161~GeV, all LEP experiments have
chosen to put limits to only one 
parameter, $\alpha_{W\Phi}$, 
%= \left(g_{ZWW} 
%-\cot\theta_W\right)\sin\theta_W\cos\theta_W$
and assume all other anomalous couplings are zero. ALEPH, DELPHI and L3 have
used only their measured total cross section and the W mass determination at
the Tevatron to derive limits, which are in the range 
$$
|\alpha_{W\Phi}| < 1.5-2.0\, ,
$$
at 95\%
confidence level. OPAL has also analyzed the kinematics of their $l\nu qq$ 
events to get a similar limit.

In the 172~GeV run, there are 
already enough events to do a full kinematic study. At
the time of this writing the four LEP experiments had all presented 
very preliminary
results using only the ``golden'' channel $l\nu qq$.
The resulting limits are
$$
|\alpha_{W\Phi}| < 0.5-1.0\, .
$$
Once the information from fully hadronic and fully leptonic events will be 
included the limit will start becoming interesting, compared to the Tevatron
ones. Already now, the L3 colaboration has combined their measurements at
161 and 172~GeV to conclude that they can exclude at 95\%
confidence level the case of no coupling between Z and W, $g_{ZWW}=0$.
Figure~\ref{fig:awf_de} shows the DELPHI measurement of the W production
angle and their fit to $\alpha_{W\Phi}$. 
\begin{figure}
\begin{center}
\place{1.0}{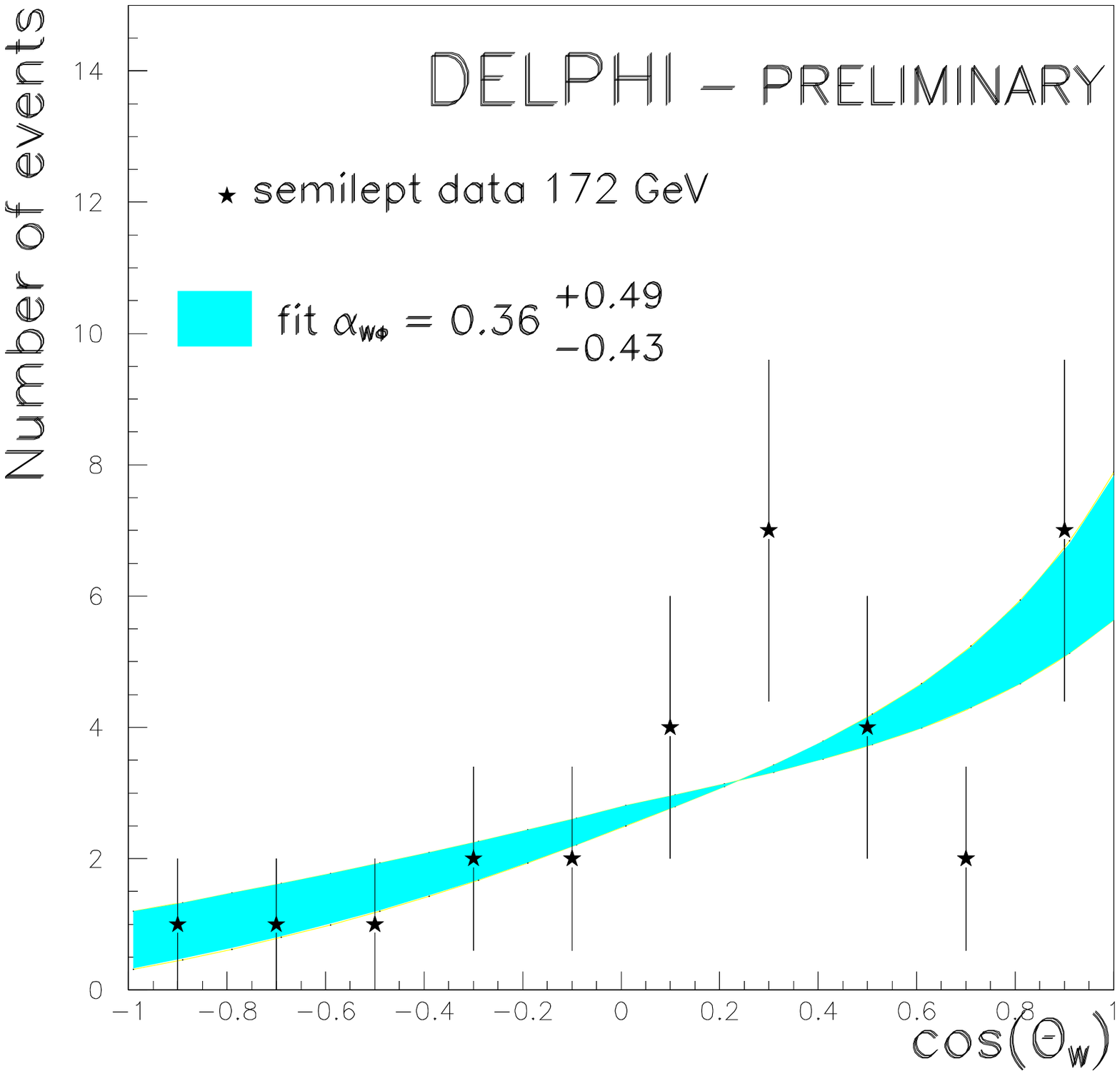}{fig:awf_de}
{Measurement of the W production angle distribution in semileptonic events
by the DELPHI collaboration and fit to $\alpha_{W\Phi}$.} 
\end{center}
\end{figure}
It is clear that this measurement will especially
benefit from higher energy and luminosity, as expected for the
1997 run.
\section{Summary}
\label{sect:summ}
The LEP2 machine has very rich capabilities in WW physics. The W mass will be
determined with total error around 30--40~MeV. And the couplings of photon and
Z to two Ws will be determined to a few per cent accuracy.

The first LEP2 run in 1996, with only a total of 21~pb$^{-1}$
of data, has produced already some very interesting results, 
like the determination 
of the W mass with 140~MeV uncertainty or the first limits 
on $\alpha_{W\Phi}$, of order 0.5.

Higher energies and luminosities in the period 1997 to 1999 or 2000 will
certainly produce more precise results, providing precise tests of the
Minimal Standard Model and, maybe, unveiling some of its weaknesses.
\section*{Acknowledgments}
It is a pleasure to thank the organizers of this meeting for their kind
invitation to give this talk and for the excellent overall organization.

\end{document}